%
%


\input harvmac.tex
\noblackbox

\newcount\yearltd\yearltd=\year\advance\yearltd by 0

\input epsf
\newcount\figno
\figno=0
\def\fig#1#2#3{
\par\begingroup\parindent=0pt\leftskip=1cm\rightskip=1cm\parindent=0pt
\baselineskip=11pt \global\advance\figno by 1 \midinsert
\epsfxsize=#3 \centerline{\epsfbox{#2}} \vskip 12pt {\bf
Figure \the\figno:} #1\par
\endinsert\endgroup\par
}
\def\figlabel#1{\xdef#1{\the\figno}}


%
\def\np#1#2#3{Nucl. Phys. {\bf B#1} (#2) #3}

\def\physrev#1#2#3{Phys. Rev. {\bf D#1} (#2) #3}

\def\jhep#1#2#3{JHEP {\bf #1} (#2) #3}
\lref\kklt{
    S. Kachru, R. Kallosh, A. Linde and S.P. Trivedi,
    ``De Sitter vacua in string theory",
    hep-th/0301240,
    \physrev{68}{2003}{046005}.
}
\lref\kst{
    S. Kachru, M. Schulz, and S. P. Trivedi,
    ``Moduli Stabilization from Fluxes in a Simple IIB Orientifold",
    hep-th/0201028.
}
\lref\gkp{
    S. B. Giddings, S. Kachru, and J. Polchinski,
    ``Hierarchies from fluxes in string compactifications",
    hep-th/0105097,
    \physrev{66}{2002}{106006}.
}
\lref\gvw{
    S. Gukov, C. Vafa and E. Witten,
    ``CFT's from Calabi-Yau four-folds",
    hep-th/9906070,
    \np{584}{2000}{69}.
}

\lref\kw{
I.~R.~Klebanov and E.~Witten,
  ``Superconformal field theory on threebranes at a Calabi-Yau  singularity,''
  Nucl.\ Phys.\ B {\bf 536}, 199 (1998)
  [arXiv:hep-th/9807080].
}
\lref\kpv{
    S. Kachru, J. Pearson and H. Verlinde,
    ``Brane/Flux Annihilation and the String Dual of a Non-Supersymmetric Field Theory",
    hep-th/0112197,
    \jhep{0206}{2002}{021}.
}
\lref\kstt{
    S. Kachru, M. Schultz, P. K. Tripathi and S. P. Trivedi,
    ``New Supersymmetric String Compactifications",
    hep-th/0211182.
}
\lref\gklm{
    S.~Gukov, S.~Kachru, X.~Liu and L.~McAllister,
    ``Heterotic moduli stabilization with fractional Chern-Simons invariants'',
    hep-th/0310159.
}
\lref\iy{
    S.~Imai and T.~Yokono,
    ``Comments on orientifold projection in the conifold and
$SO\times USp$ duality cascade'',
    hep-th/0110209,
    \physrev{65}{2002}{066007}.
}
\lref\dbchem{
C.~P.~Burgess, N.~E.~Grandi, F.~Quevedo and R.~Rabadan,
  ``D-brane chemistry,''
  JHEP {\bf 0401}, 067 (2004)
  [arXiv:hep-th/0310010].
}
\lref\smallna{
    S.~Dimopoulos, S.~Kachru, N.~Kaloper, A.~E.~Lawrence and E.~Silverstein,
    ``Small numbers from tunneling between brane throats'',
    hep-th/0104239,
    Phys.\ Rev.\ D {\bf 64}, 121702 (2001).
}
\lref\smallnb{
    S.~Dimopoulos, S.~Kachru, N.~Kaloper, A.~E.~Lawrence and E.~Silverstein,
    ``Generating small numbers by tunneling in multi-throat compactifications'',
    hep-th/0106128.
}
\lref\comcon{
    P.~Candelas and X.~C.~de la Ossa,
    ``Comments On Conifolds'',
    Nucl.\ Phys.\ B {\bf 342}, 246 (1990).
}
\lref\cgqu{
    J.~F.~G.~Cascales, M.~P.~G.~del Moral, F.~Quevedo and A.~Uranga,
    ``Realistic D-Brane Models on Warped Throats: Fluxes, Hierarchies and Moduli Stabilization'',
    hep-th/0312051.
}
\lref\ceresole{
    A.~Ceresole, G.~Dall'Agata, R.~D'Auria and S.~Ferrara,
    ``Spectrum of type IIB supergravity on $AdS_5 \times T^{1,1}$ :
Predictions on $N  = 1$ SCFT's'',
    hep-th/9905226,
    Phys.\ Rev.\ D {\bf 61}, 066001 (2000).
}
\lref\ceresoleb{
    A.~Ceresole, G.~Dall'Agata and R.~D'Auria,
    ``KK spectroscopy of type IIB supergravity on $AdS_5\times T^{1,1}$'',
    hep-th/9907216,
    JHEP {\bf 9911}, 009 (1999).
}
\lref\rsii{
    L.~Randall and R.~Sundrum,
    ``A large mass hierarchy from a small extra dimension'',
    Phys.\ Rev.\ Lett.\  {\bf 83}, 3370 (1999) [arXiv:hep-ph/9905221].
}
\lref\rsi{
    L.~Randall and R.~Sundrum,
    ``An alternative to compactification'',
    Phys.\ Rev.\ Lett.\  {\bf 83}, 4690 (1999) [arXiv:hep-th/9906064].
}
\lref\ks{ I.~R.~Klebanov and M.~J.~Strassler,
    ``Supergravity and a confining gauge theory: Duality cascades and chiSB-resolution of naked singularities'',
    JHEP {\bf 0008}, 052 (2000) [arXiv:hep-th/0007191].
}

\lref\VerlindeFY{
  H.~L.~Verlinde,
  ``Holography and compactification,''
  Nucl.\ Phys.\ B {\bf 580}, 264 (2000)
  [arXiv:hep-th/9906182].
}

\lref\DeWolfeQX{
  O.~DeWolfe, S.~Kachru and H.~L.~Verlinde,
  ``The giant inflaton,''
  JHEP {\bf 0405}, 017 (2004)
  [arXiv:hep-th/0403123].
}

\lref\denef{
  F.~Denef,
  ``Constructions and distributions of flux vacua,''
  talk given at Strings 2005 conference,
  {\tt http://www.fields.utoronto.ca/audio/05-06/strings/denef/}
}

\def\CM{{\cal M}}
\def\IC{\relax{\rm I\kern-.54em C}}
\def\IZ{\relax\ifmmode\hbox{Z\kern-.4em Z}\else{Z\kern-.4em Z}\fi}
\def\IR{\relax{\rm I\kern-.18em R}}
\def\II{\relax{\rm 1\kern-.25em I}}

\def\CL{{\cal L}}

\def\frac#1#2{{#1 \over #2}}

\def\det{{\rm det}}
\def\tr{{\rm tr}}
\def\Tr{{\rm Tr}}
\def\Im{{\rm Im}}

\def\dbar{\overline {\rm D}}

\def\dbar#1{\overline{{\rm D}#1}}

\def\exp#1{{e}^{#1}}

\Title{
\vbox{\baselineskip12pt\hbox{\tt hep-th/0508080}
\hbox{WIS/20/05-AUG-DPP}
}}
{\vbox{ {\centerline{Open String Moduli in KKLT Compactifications}}
}}
\bigskip
\centerline{
    Ofer Aharony$^{}$\foot{E-mail: {\tt Ofer.Aharony@weizmann.ac.il}.},
    Yaron E. Antebi$^{}$\foot{E-mail: {\tt ayaron@weizmann.ac.il}.},
    and Micha Berkooz$^{}$\foot{E-mail: {\tt Micha.Berkooz@weizmann.ac.il}.}
}
\bigskip
\centerline{{\it Department of Particle Physics,}}
\centerline{{\it The Weizmann Institute of Science, Rehovot
76100, Israel}}
\bigskip
\medskip
\noindent

In the Kachru-Kallosh-Linde-Trivedi (KKLT) 
de-Sitter construction one introduces an
anti-D3-brane that breaks the supersymmetry and leads to a positive
cosmological constant. In this paper we investigate the open string
moduli associated with this anti-D3-brane, corresponding to
 its position on the $S^3$ at the tip of the deformed conifold. We
show that in the KKLT construction these moduli are very light, and we
suggest a possible way to give these moduli a large mass by putting
orientifold planes in the KKLT ``throat".

\Date{August 2005}

\newsec{Introduction and Summary}

Type II string compactifications to four spacetime dimensions with
non trivial RR and NSNS background fluxes have been studied
extensively in the literature in the past few years, as a way to
stabilize moduli in string theory.
Compactifications on generic Calabi-Yau three-folds without background fluxes 
lead to hundreds of massless scalar
moduli fields, causing various phenomenological problems since no
light scalar fields have been observed in nature.
However, by turning on some background value for the fluxes on cycles
of the Calabi-Yau, a potential develops that
stabilizes those moduli at some fixed value and generates a mass for
the scalar fields (see \denef\ and references therein).

Several examples of this mechanism, involving orientifolds, have been
studied in detail.
In type IIA string theory
there are several known examples of
toroidal orientifolds in which all moduli are
stabilized.
In type IIB string theory, the classical supergravity action generates a
potential for the complex structure moduli of the Calabi-Yau manifold but
not for its K\"ahler structure moduli.
Since the total volume
of the compact manifold is a K\"ahler modulus, it is not possible to fix
all moduli by fluxes in the type IIB supergravity approximation.
However, it has been argued \kklt\ that
non-perturbative effects in type IIB string theory,
such as gauge theory instantons or gaugino condensation
in the worldvolume of D7-branes
or wrapped Euclidean D3-branes,
generate a potential which depends also on the K\"ahler moduli.
Including these nonperturbative
effects leads to a potential with a minimum with a negative cosmological
constant,
describing a supersymmetric Anti de-Sitter (AdS) background.

The authors of \kklt\ suggested that a slight modification of such a
background could lead to a meta-stable de Sitter (dS) background, in
agreement with recent observations suggesting a positive
cosmological constant. The modification involves introducing a
space-filling anti-D3-brane (which we will denote as a $\dbar3$-brane)
which raises the potential energy. This breaks
all the supersymmetry, and using some fine tuning it was argued that it is
possible to obtain a positive yet small cosmological constant. Following
the work of \kklt, various other suggestions for constructing meta-stable
dS vacua have also appeared.

In addition to changing the potential, the addition of the $\dbar3$ brane
has implications regarding the moduli in the theory. In the presence of
the $\dbar3$ brane there is also an
open string sector, which includes some light scalar fields (moduli)
that can be
interpreted as the location of the $\dbar3$ brane in the compact space.
In this paper we study these moduli.

We begin in section 2 by reviewing the KKLT construction, in which the
moduli are stabilized near a conifold singularity such that the
compactification includes a Klebanov-Strassler (KS) \ks\ type ``throat'',
generating a hierarchy by a factor of the small warp factor $a_0$ at the
tip of the ``throat'' \gkp, and
a $\dbar3$ brane is then added at the tip of the ``throat''.
In section 3 we discuss the mass of the open string moduli corresponding
to the position of the $\dbar3$ brane. We argue that in the limit of an
infinite ``throat'' these moduli are massless since they are Goldstone
bosons, but when the ``throat'' is finite the background is changed and
the moduli obtain a mass. We discuss in detail the deviation of the
finite ``throat'' theory from the infinite ``throat'' theory of \ks, and
we identify the leading deviation which contributes to the mass of the
open string moduli. We use the approximate conformal symmetry of the
``throat'' theory to classify the deviations, and we find that the leading
deviation corresponds to an operator of dimension $\Delta=\sqrt{28}\simeq
5.29$, and that it leads to a mass squared for the open string moduli scaling
as $a_0^{\Delta-2} \simeq a_0^{3.29}$. In the interesting limit of large
warping, $a_0 \ll 1$, this mass is exponentially
lighter than the other mass scales
appearing in the warped compactification, implying that the KKLT scenario
generally leads to light scalars which could cause phenomenological problems.

In section 4 we suggest a possible way to resolve this problem and
increase the mass of the moduli, by positioning two of the orientifold
3-planes (which must be present anyway in KKLT-type compactifications)
at the tip of the ``throat'', and adding to them half-D3-branes so
that they become ${\rm O}3^+$-planes rather than ${\rm O}3^-$
planes. The $\dbar3$ brane is then attracted to these ${\rm O}3^+$
planes, increasing the mass of the open string moduli. The mass
squared is still smaller than the typical mass scales, but only by a
factor of the string coupling $g_s$ which does not have to be very
small, so this may not lead to phenomenological problems (especially
if the standard model fields live in a different position in the
Calabi-Yau and couple very weakly to the $\dbar3$ brane fields). Our
scenario has the added advantage that by adding two half D3-branes in
addition to the $\dbar3$ brane we do not generate a tadpole for the
D3-brane charge, unlike the original KKLT scenario where such a
tadpole exists and leads to subtleties in using the probe
approximation for describing the $\dbar3$ brane (due to the necessity
to change the background elsewhere to compensate for the $\dbar3$
brane charge).

Finally, in two appendices we derive some results used in the text. In
appendix A we list the possible deformations of the $AdS_5\times T^{1,1}$
background (which is a good approximation to the ``throat'') which can appear
as deformations of the ``throat'' in our background. In appendix
B we discuss the moduli space of the gauge theory dual to the ``throat''
region after deformations by superpotential operators, and we argue that any
such deformations reduce the dimension of the moduli space.

\newsec{A review of dS flux compactifications with $\dbar{3}$-branes}

The setting for our analysis in the following sections is the dS
background of KKLT \kklt. We start with a brief overview of a general
flux compactification and then proceed to describe the construction
of the dS background. More details can be found in
\refs{\kklt,\gvw,\gkp}.

\subsec{Warped flux compactifications}

We consider type IIB string theory in the supergravity
approximation, described in the Einstein frame by the action
\eqn\IIBaction{\eqalign{
    S_{IIB}=&
    \frac{1}{2\kappa_{10}^2} \int\! d^{10}x \sqrt{-g}\left\{
    \CR-\frac{\partial_M\tau\partial^M\bar\tau}{2(\Im\tau)^2}
    -\frac{G_3\cdot\bar G_3}{12\Im\tau}
    -\frac{\tilde F^2_5}{4\cdot5!}
    \right\}\cr
    &+\frac{1}{8i\kappa_{10}^2}\int
    \frac{C_4\wedge G_3\wedge G_3}{\Im\tau}
    +S_{local},
    }}
where $\tau=C_0+i\exp{-\phi}$ is the axio-dilaton field and we
combine the RR and NS-NS three-form fields into the generalized complex
three-form field $G_3=F_3-\tau H_3$. In addition one must impose a self
duality condition on the five form $\tilde F_5 \equiv F_5-\frac12 C_2\wedge
H_3 +\frac12 B_2\wedge F_3$,
\eqn\selfdual{
    \tilde F_5=*\tilde F_5.
    }
The local action $S_{local}$ includes the contributions from
additional local objects such as D-branes or orientifold planes.

We begin by considering warped backgrounds, with a metric
of the form
\eqn\wmetric{
    ds_{10}^2={\rm e}^{2A(y)}\eta_{\mu\nu}dx^\mu dx^\nu +
    {\rm e}^{-2A(y)}\tilde g_{mn}(y)dy^m dy^n,
    }
where $\mu,\nu=0,1,2,3;m,n=4,\cdots,9$,
and the unwarped metric
$\tilde g_{mn}$ scales as $\sigma^{1/2}$, where $\sigma$ is the imaginary
component of the complex K\"ahler modulus related to the overall
scale of the compact Calabi-Yau. In addition, both the five form and
three form fields are turned on. Due to $4$-dimensional Poincar\'e
invariance only compact components of $G_3$ may be turned on, while
for the five form, the Bianchi identity determines it to be of the
form
\eqn\fiveform{
    \tilde F_5=(1+*)d\alpha(y)\wedge dx^0\wedge dx^1\wedge dx^2\wedge
    dx^3.
    }
Finally, local objects extended in the four non-compact dimensions can be added
wrapping cycles of the compact space. These must satisfy the tadpole
cancellation condition
\eqn\tadpole{
    \frac{1}{2\kappa_{10}^2T_3}\int_{\CM_6}H_3\wedge F_3
    +Q_3^{local}=0,
    }
where $Q_3^{local}$ is the D3-brane charge of the local objects.

The supergravity equations of motion for such a configuration of fields
can be conveniently written in terms of the following combinations of
the five form and warp factor
\eqn\fieldcomb{
    \Phi_\pm\equiv\exp{4A}\pm\alpha.
    }
The Einstein equation and the  Bianchi identity for the 5-form field can be
combined to give
\eqn\phieom{
    \tilde\nabla^2\Phi_{\pm}=\frac{\exp{2A}}{6\Im\tau}|G_{\pm}|^2
    +\exp{-6A}|\nabla\Phi_{\pm}|^2
    +local,
    }
where we defined the imaginary self dual (ISD) and imaginary anti
self dual (IASD) components of the generalized three form flux,
\eqn\isdparts{
    G_\pm=iG\pm*_6G \qquad\Rightarrow\qquad *_6G_\pm=\pm iG_\pm.
    }
The local objects act as sources for the fields $\Phi_\pm$. D3-branes
and O-planes appear as sources only in the equation for $\Phi_+$,
while $\dbar3$-branes appear only in the equation for $\Phi_-$. For a
background with no $\dbar3$-branes there are no sources for $\Phi_-$, so
we get using \phieom\ and the compactness of the Calabi-Yau
\eqn\warpsoln{
    \Phi_-=0 \Rightarrow \alpha=\exp{4A}.
    }
Since $|G_-|^2$ is positive definite it must vanish everywhere and so
$G_3$ is ISD.

The equations of motion can be compactly summarized by a
$4$-dimensional superpotential \gvw
\eqn\spot{
    W=\int \Omega\wedge G_3,
    }
where $\Omega$ is the holomorphic $(3,0)$ form, together with the
standard supergravity K\"ahler potential. This notation makes
explicit the fact that the equations give non-trivial restrictions on
some of the moduli. The superpotential depends both on the axio-dilaton
(through its appearance in $G_3$) and on the geometrical complex
structure moduli that appear in $\Omega$. However, the resulting
four dimensional supergravity theory is of
the no scale class. The K\"ahler moduli, including the global
volume of the compact manifold, have no potential (in the supergravity
approximation) and remain unfixed.

Consider probing this space with D$3$-branes. The D3-brane action in the
Einstein frame without turning on any open string fields,
including both the DBI and the Wess-Zumino term, is given by
\eqn\dactn{
    S_{{\rm D}3}=-T_3\int\sqrt{g_4}d^4x \Phi_-.
    }
For the type of solutions discussed above, obeying equation \warpsoln, we
obtain that these probes feel no force, and their moduli space is the full
compact manifold. For $\overline {{\rm D}3}$-brane probes,
due to the opposite sign in the Wess-Zumino term, we find that the action is
\eqn\bardactn{
    S_{\overline{{\rm D}3}}=-T_3\int\sqrt{g_4}d^4x \Phi_+.
    }
In our background where $\Phi_+=2\exp{4A}$ there is thus  a force on the
$\dbar3$-brane driving it towards smaller values of the warp factor.

\subsec{Getting a hierarchy from the conifold}

It is phenomenologically interesting to find a background in which,
in addition to fixing the moduli, there is a large warped throat.
This can be used to realize the construction of Randall and Sundrum
\refs{\rsi,\rsii,\VerlindeFY},
giving a solution to the hierarchy problem. Such a
background was found in \gkp\ by considering a generic Calabi-Yau
near a special point in its moduli space where it develops a
singularity. Generically such a singularity looks locally
like the conifold singularity \comcon\ which can be described by the
sub-manifold of $\IC^4$ defined by:
    \eqn\conifold{
        z_1^2+z_2^2+z_3^2+z_4^2=0.
    }
The conifold is a cone whose base is $T^{1,1}=(SU(2)\times SU(2))/U(1)$,
a fibration of $S^3$
over $S^2$. The cone is singular at $(z_1,z_2,z_3,z_4)=(0,0,0,0)$
where the spheres shrink to zero size. The isometry group of
the base geometry is easily seen to be $SU(2)\times SU(2)\times
U(1)$, where the $SU(2)\times SU(2)\simeq SO(4)$ rotates the $z_i$'s
and the $U(1)$ adds a constant phase $z_i\to \exp{i\alpha}z_i$.

The singularity of the conifold can be smoothed in two ways, by
blowing up either of the spheres to a finite size. We will be
interested in the deformation of the conifold, which is the
sub-manifold given by
    \eqn\defcon{
        z_1^2+z_2^2+z_3^2+z_4^2=\mu,
    }
where $\mu$ becomes a complex structure modulus for this manifold.
Geometrically, in \defcon\ the $S^2$ shrinks to zero size at the tip while the
$S^3$ remains at some finite size. The minimal size $S^3$ at the ``tip"
 is given by
    \eqn\contip{
        |z_1|^2+|z_2|^2+|z_3|^2+|z_4|^2=|\mu|.
    }
This deformation breaks the symmetry group to $SU(2)\times
SU(2)\times\IZ_2$, where the $SO(4)$ can be understood geometrically
as rotations of the $S^3$.

Placing $M$ fractional D3-branes at a conifold singularity, the
background near the singularity is given, for large $g_sM$ and for some
range of radial distances from the singularity, by the
KS solution \ks, where in the near horizon geometry
one replaces the branes by fluxes. Such a configuration involves
turning on $M$ units of $F_3$ flux on the $S^3$ at the tip of the
conifold, and also ($-K$) units of $H_3$ flux on
the dual cycle (which is non-compact in \ks\ but is compact when
we embed this into a compact Calabi-Yau).
It is customary to define $N=MK$. In \gkp\ it
was found that such fluxes generate a warped throat similar to \ks\ near the
singularity. The superpotential stabilizes the complex structure
modulus $\mu$ at a value for which the warp factor at the tip of the
throat is given by
    \eqn\wfactor{
        a_0\equiv \exp{A_0}=\exp{-2\pi K/3Mg_s},
    }
which is exponentially small when $K\gg g_sM$ (the validity of the supergravity
approximation in the ``throat'' requires also $g_s M \gg 1$).

In the throat region of the Calabi-Yau the warp factor is given by
\eqn\thraotwarp{
    \exp{-4A}=\frac{27\pi}{4u^4}\alpha'^2g_sN
        \left(1+\frac{g_sM}{K}
            \left(\frac{3}{8\pi }+\frac{3}{2\pi }\ln(\frac{u}{u_0})
            \right)
        \right)
    }
where $u$ is the radial coordinate along the throat. At the tip of
the throat the redshift is minimal and given by \wfactor. There we
get $u\sim Ra_0$, where we defined $R^4=\frac{27}{4}\pi\alpha'^2 g_s N$.
The bulk of the Calabi-Yau, where the warp factor is of order unity
(and deviations from \thraotwarp\ are large)
is at $u\sim R$.

\subsec{Lifting to a dS background}

Although phenomenologically interesting, backgrounds of this type classically
have
at least one scalar modulus. The low-energy theories we arrive at are
no-scale models, and the potential generated by the fluxes does not give
any mass term for the K\"ahler modulus related to the volume
of the compact space. This was mended in \kklt\ by considering
non-perturbative effects. Terms in the potential coming either from
instantons in non-Abelian gauge groups on a stack of D$7$-branes or
from Euclidean D$3$-branes wrapped on 4-cycles
depend on the volume of the space, and
stabilize it at some finite value.

The stabilization of the K\"ahler modulus leads to a vacuum with a
negative cosmological constant, an AdS space. It was then argued that
adding an $\overline {{\rm D}3}$-brane (which, as discussed above,
should sit at the ``tip" of the throat to be stable) results in a positive
contribution to the scalar potential from \bardactn,
and with some tuning of the
parameters it can lift the minimum of the potential to a small
positive value. Thus it is possible to get a de-Sitter space with
a small cosmological constant.

\newsec{The $\overline{{\rm D}3}$-brane moduli}

A consequence of the introduction of an $\dbar{3}$-brane to the
warped background is the addition of new light scalar fields from the
open strings ending on the $\dbar3$-brane, corresponding to the
position of the $\dbar{3}$-brane on the compact space. In this
section we analyze the potential for these moduli in the KKLT
background. We first consider the background without the additional
$\dbar3$-brane, and estimate the deviation of the warped background
with a compact Calabi-Yau from the non-compact background of \ks.
We then use this to
estimate the masses for the position of the $\dbar3$-brane, using the action
\bardactn\ and considering the $\dbar3$-brane as a probe (as in \kklt).
This approximation is valid when $g_s \ll 1 \ll g_sM$.

From \bardactn\ we see that the $\dbar3$-branes are not free to move
on the compact space since they have a non trivial potential
proportional to the warp factor. This potential drives them to the
tip of the throat where the warp factor is minimal, giving a mass to
the scalar field corresponding to the radial position of the $\dbar3$-brane.

At the tip of the throat, the $\dbar3$-brane can still move on the
$S^3$. In the full infinite KS solution there is an
exact $SO(4)$ symmetry corresponding to rotations in this 3-sphere,
and placing the $\dbar3$-brane breaks this symmetry as $SO(4)\to
SO(3)$. This gives rise to three massless moduli, the three Goldstone bosons,
which can also be interpreted as the three coordinates of the
position of the $\dbar3$-brane in the $S^3$.

In our background
there are, however, corrections coming from the compactness of the
Calabi-Yau, as the background deviates from the KS
solution away from the tip. From the point of view of the field theory dual
of the KS background, these corrections are related to UV
perturbations (changes in coupling constants). Some of these corrections
explicitly break the $SO(4)$ symmetry, and thus generate a mass for the
Goldstone bosons. We will
first classify the possible perturbations that can be turned
on in this class of backgrounds, and then go on to consider their
effect on the mass for the three moduli of the $\dbar3$-brane.

\subsec{UV corrections of the background}

The deformation of our background away from the KS geometry,
at large radial position away from
the conifold,
is easily described in the language of the dual field
theory. The dual theory (at some cutoff scale) has an $SU(N)\times
SU(N+M)$ gauge group, with gauge superfields $W_1$ and $W_2$ corresponding
to the two gauge groups, and two doublets of chiral superfields
$A_i,B_i$ ($i=1,2$) in the $(N,\overline{N+M})$ and $(\overline
N,N+M)$ representations, respectively, of the $SU(N)\times SU(N+M)$
group, and in the $(\frac12,0)$ and $(0,\frac12)$ representations of the
global $SU(2)\times SU(2)$ symmetry.

In the dual description the region near the singularity describes the
low-energy physics of the field theory while the Calabi-Yau end
of the throat serves as a UV cutoff of the field theory. Deforming the
solutions at large radial position is described by changing the theory
at some large UV scale where the effective theory is some deformation
of the KS theory,
\eqn\deflag{
    \CL=\CL_{KS}+c_i\int\CO_i.}
Generally all possible operators might be turned on at this scale,
and they could influence the $\dbar3$-brane at the tip (the IR
limit) and give a mass to the moduli. Due to the renormalization group flow
the contribution to the mass of the $\dbar3$-brane at the tip
will be dominated by the most relevant operators at the IR, namely
the lowest dimension operators. Relevant and marginal operators will
have a large effect, while that of the irrelevant operators will be
suppressed.

It is sufficient to
analyze the operators and their dimensions in the conformal
case \kw\ where the gauge group is $SU(N)\times SU(N)$, since the
cascading case is expected to behave similarly up to log
corrections and operator mixings which should not change our
conclusions. For this case the classification of all supergravity
KK-modes on $T^{1,1}$ and the corresponding operators in the field
theory was given in \refs{\ceresole,\ceresoleb}. Since we are only
interested in turning on operators that break neither 4-dimensional
Lorentz invariance nor supersymmetry, we can restrict our attention
to the highest components of the different superconformal multiplets
and consider only those that are Lorentz scalars. The only possible
operators come from vector multiplets of the five dimensional gauged
supergravity which
arises by KK reduction on $T^{1,1}$,
either long multiplets or chiral multiplets.

The analysis of supergravity modes is carried out in appendix A,
where we find only one possible relevant operator
\eqn\relop{
    S_1=\int\!\! d^2\theta\ \Tr(A^iB^j),\qquad i,j=1,2,\qquad
    \Delta_{S_1}=2.5,
} and three possible marginal operators
\eqn\margop{\eqalign{
    S_2&=\int\!\! d^2\theta\ \Tr(A^iB^jA^kB^l),\qquad \Delta_{S_2}=4,\cr
    \Phi_0&=\int\!\! d^2\theta\ \Tr(W_1^2+W_2^2),\qquad \Delta_{\Phi_0}=4,\cr
    \Psi_0&=\int\!\! d^2\theta\ \Tr(W_1^2-W_2^2),\qquad \Delta_{\Psi_0}=4.
}} 
The operator $S_2$ is symmetric in $(i,k)$ and $(j,l)$; the anti-symmetric
combination mixes with $\Phi_0$.
There is also an infinite number of irrelevant operators, all of them
with dimensions $\Delta\ge5.29$.

In fact not all possible operators are turned on in the compact
Calabi-Yau background. As discussed above,
a probe D-brane in this background must feel no force and its moduli
space should describe the full 6 dimensional compact geometry. In appendix
B it is found that the addition of the operators $S_1, S_2$ changes
the moduli space drastically and necessarily results in a force on
the D$3$-brane. Thus, these operators are not turned on in the warped flux
compactifications.

The two marginal operators, $\Phi_0$ and $\Psi_0$, can be turned on,
but they are symmetric under the $SU(2)\times SU(2)$ and do not
lead to symmetry breaking and to a mass for the $\dbar3$-brane
moduli. From the field theory perspective they correspond to changing
the coupling constants that are already present in the non-deformed theory
and do not generate new terms in the action.

\subsec{Masses from UV corrections}

In the previous subsection we have seen that relevant operators
are not turned on in the warped background, while the possible marginal
operators do not break the symmetry and leave the moduli massless.
Irrelevant operators, however, can be turned on, and we next discuss the masses
generated by them. As discussed in
appendix A, the various operators which preserve SUSY and Lorentz
invariance are related to Kaluza-Klein modes of the warped
metric on the $T^{1,1}$ $\hat g_{ij}$, the field $\Phi_+$ defined in
\fieldcomb, the three form field $G_3$ and the axio-dilaton $\tau$.

The operators are turned on at the UV cutoff, and in order to consider
their effect on the IR physics we need to discuss their flow, or in
the supergravity language their profile along the radial
coordinate. We start by considering the profiles of the fields corresponding
to these operators on $AdS_5\times
T^{1,1}$, using the metric $ds_{AdS}^2 = u^2 dx^{\mu} dx_{\mu} + du^2 / u^2$.
There are two independent solutions for the field $\phi$ corresponding
to an operator of dimension $\Delta>2$, with the following $u$-dependence :
\eqn\twomodes{
    \phi(u) = au^{-\Delta}+bu^{\Delta-4}.
    }
In the KS background there are small logarithmic corrections to this, and
in addition the behavior near the tip of the ``throat'' gives some IR
boundary condition for the field equations. Generically this implies
that at $u=Ra_0$
the two terms are of the same order. Then, at the UV cutoff $u \sim R$
(the Calabi-Yau), the
second term will dominate so
\eqn\atcy{
    \phi(u\sim R) \simeq bR^{\Delta-4}.
    }
Deforming the theory at the UV by some $\delta\phi(R)\sim\phi_0$ will
then correspond in the IR to
\eqn\irdef{
    \delta\phi(Ra_0) \sim
    \phi_0\frac{(Ra_0)^{\Delta-4}}{R^{\Delta-4}}
    =\phi_0a_0^{\Delta-4}.
    }
The deformation in the IR is suppressed for operators with higher
dimension, as expected.

The largest contribution to a mass of an object localized near the tip
will be from the
operator with lowest dimension that breaks the $SU(2)\times
SU(2)$ global symmetry. The analysis of the previous subsection and
Appendix A implies that this is the lowest component of
the vector multiplet I, with $j=l=1$ and $r=0$, whose dimension is
$\Delta=\sqrt{28}\simeq5.29$. This operator corresponds in the
supergravity to a KK mode of the warped metric $\hat g_{ij}$. We do
not see any reason why this operator should not appear in the CY
compactification so we assume
that it does\foot{Note that this operator deforming the throat seems to be
different from the one analyzed in the appendix of \DeWolfeQX.}.
At the UV we have $\hat g_{ij}\sim\sigma^{1/2}$, and we expect
the deformation of the metric to be of the same order as the
metric so we can approximate $\delta \hat g_{ij}|_{\rm UV}\sim\sigma^{1/2}$
and
\eqn\gdefnew{
    \delta \hat g_{ij}|_{\rm IR}\sim\sigma^{1/2} a_0^{\Delta-4}
    =\sigma^{1/2} a_0^{1.29}.
    }

In order to evaluate the corresponding mass we need to write in more
detail the action on a probe $\dbar3$-brane. We consider a
$\dbar3$-brane filling the non compact space-time and positioned at
the point $X^m$ in the compact Calabi-Yau, which is near the tip of
the throat. Expanding around this position, we get the action
\bardactn\ with an additional kinetic term. Using the solution of the
supergravity equations for the five form
\warpsoln\ we get
\eqn\dbaractnkin{
    S_{\overline{{\rm D}3}}
    =-2T_3\int\sqrt{g_4}d^4x \exp{4A(X^m)}
    -T_3 \alpha'
\int\sqrt{g_4}d^4x \frac12 g_4^{\mu\nu}\partial_\mu X^m \partial_\nu X^n \tilde
    g_{mn},
    }
where $\tilde g_{mn}$ is the unwarped metric, $\tilde g_{mn}\simeq a_0^2\hat
g_{mn}$. Mass terms appear in this action only through the
dependence of the warp factor, $\exp{4A}$, on the position $X^m$.
Since in the full non-compact case the warp
factor has only radial dependence, for a mass to be generated in the $S^3$
directions we need
to consider the change in the warp factor due to the deformed
supergravity fields $\hat g_{ij}$.

Tracing the Einstein equation and using \warpsoln\ we can write the
equation of motion for the warp factor as
\eqn\Aeom{
    \hat\nabla^2A
    =\frac{g_s}{48}|G|^2,
    }
where $\hat\nabla^2$ is the Laplacian on the warped compact space and
we use the warped metric to raise and lower indices. The change in
$A$ due to the deformation in $\hat g_{ij}$ will thus satisfy
\eqn\Aeom{
    \hat\nabla^2\delta A
    =\frac{g_s}{48}G_{m_1n_1i}G^*_{m_2n_2j}\hat g^{m_1m_2}
\hat g^{n_1n_2}\delta\hat g^{ij}
    \sim a_0^{\Delta-4}.
    }
In this equation we dropped a term $(\delta\hat\nabla^2) A$ since the
change in the Laplacian due to the deformation in the compact metric
$\hat g_{ij}$
will be proportional to derivatives in those directions, while the
original $A$ has no such dependence and so this term vanishes.

The masses arise due to the variation of $A$ in the $S^3$ at the tip,
where the $\dbar3$-brane position is
parameterized by $X^i$. Since in the warped metric we are using, this
3-sphere has constant size (with a radius $\sim \sqrt{g_s M}$), we can estimate
\eqn\aonsphere{
    A\sim A_0+(g_s M)^{-1} a_0^{\Delta-4} \hat g_{ij} X^iX^j.
    }
Plugging into \dbaractnkin\ we find
\eqn\dbaractnkin{
    S_{\overline{{\rm D}3}} \sim -T_3\int\sqrt{g_4}d^4x\left[
    2a_0^4
    +2(g_s M)^{-1} a_0^{\Delta-2} {\tilde g}_{ij} X^iX^j
    +\frac{\alpha'}{2} g_4^{\mu\nu}\partial_\mu X^m \partial_\nu X^n \tilde g_{mn}\right]
    }
where we changed the metric to the unwarped metric in both kinetic
and mass terms\foot{Note that the first term, even though naively it is
independent of the volume $\sigma$ of the compact space, actually does
give a potential for the volume factor \kklt\ when we rescale the four
dimensional metric canonically \kpv.}.
We see that a mass term was generated with a mass of
the order of $m^2\sim (g_s M \alpha')^{-1} a_0^{\Delta-2}=(g_s M \alpha')^{-1}
a_0^{3.29}$.

We see that the deformation of the theory at the UV does indeed
generate mass terms for the open string moduli. However, the highest
contribution is of order $a_0^{3.29}$. In the warped background
the typical IR mass scale is of order $a_0^2$,
so the mass generated here is exponentially smaller (given \wfactor). In the
Klebanov-Strassler background there are presumably subleading
logarithmic corrections to this result, however it is still highly
suppressed.
Such light moduli would lead to phenomenological problems if we
try to use such a scenario to describe the real world.

\newsec{A large open string moduli mass from O-planes}

It is possible to obtain a higher mass for the open string moduli by
using O-planes. In this section we calculate this mass.
Recall that integrating the supergravity equation of motion \phieom\ on the
compact space we get that the left hand side vanishes since there are
no boundaries. The right hand side is positive definite, except for possible
negative contributions in the local terms corresponding to
orientifold planes. Hence in general we must have orientifold
3-planes in order to be able to solve the equations of motion. It is
then natural to try and use these orientifolds for the purpose of
stabilizing the moduli for the $\dbar3$-brane, by choosing the position
of these orientifolds to be at the tip of the throat.

For simplicity we consider an orientifold of the non-compact
Klebanov-Strassler solution. Since the analysis is local, embedding
this into the full background will not change the conclusions.
The action of the orientifold is defined
as in \iy\ by
    \eqn\orienti{
        (z_1,z_2,z_3,z_4) \to (z_1,-z_2,-z_3,-z_4).
    }
This orientifold has two fixed points, both on the tip of the deformed
conifold \contip\ at the poles of the $S^3$, $(z_1,z_2,z_3,z_4)=
(\pm\sqrt{\mu},0,0,0)$.
Physically there are two ${\rm O}3$-planes at these points, which will
interact with the $\dbar3$-brane and generate a potential for its
position on the 3-sphere.
Note that the addition of the ${\rm O}3$-planes
has no effect on the supersymmetry of
the model, since the ${\rm O}3$-planes break
the same supercharges as the fluxes.

The D$3$ charge of an orientifold plane as well as its tension is
negative (equal to $-1/4$ that of a D3-brane),
while for $\dbar3$-branes the charge is negative and the
tension is positive. We see that both effects result in a repulsive
force, so that the $\dbar3$-brane does not get stabilized but rather
it would want to sit on the equator of the $S^3$. However one
can use half-D$3$-branes to fix the situation. Putting a
half-D3-brane on the orientifold singularity, we get an ${\rm
O}3^+$-plane with the opposite charge and tension. Since we have two
singular points we can add two such half-D3-branes to make both O-planes
positively charged. Note that the insertion of one additional unit of
D$3$-brane charge is actually a
necessity once we introduce the $\dbar3$-brane,
due to the tadpole cancellation condition. Assuming that
without $\dbar3$-branes the background with ${\rm O}3^-$ planes
is a solution of the
supergravity, inserting the $\dbar3$-brane will cause a deficiency in
D$3$ charge, which can be resolved by the extra two half D3-branes.
Note that an $\dbar3$-brane cannot annihilate with a half-D3 so the solution
should still be (meta)-stable.

The potential between the $\dbar3$-brane and the ${\rm O}3^+$-plane can
be calculated by worldsheet methods \dbchem.
The first contribution which depends on the distance between the
$\dbar3$-brane and the orientifold comes from the M\"obius strip, and
is equal to
\eqn\mobius{
        \CM=2T_3g_s \sum_{n=0}^{\infty} c_n r^{2n},
    }
where $r$ is the distance (in string units) between the $\dbar3$-brane and the
${\rm O}3^+$-plane, namely
$r^2=\hat g_{ij} X^iX^j$, and the coefficients are
\eqn\coeffsa{
        c_n=(-1)^{n+1}k_{n-3}\left(\frac{2^{n-4}\pi^{(2-n)/2}}{n!}\right),
    }
where $k_{n-3}$ is a positive number for $n=0,1$ \dbchem.
This computation was done in flat space, but for large $g_s M$ the
curvature is small and it is a good approximation.

The first term in the expansion is a correction to the energy which is
independent of the distance.
The second term is a
quadratic potential for the position of the $\dbar3$-brane which
describes attraction between the $\dbar3$-brane and the ${\rm O}3^+$-plane.
The contribution to the action is
\eqn\actndiskcap{
        -2T_3\int\sqrt{g_4}d^4x a_0^4
        g_s c_1 \hat g_{ij} X^iX^j
        = -2T_3\int\sqrt{g_4}d^4x a_0^2
        g_s c_1 {\tilde g}_{ij} X^iX^j,
        }
so that the mass of the $X^i$ fields is
\eqn\mmass{
        m^2\sim \frac{2 c_1}{\alpha'}g_s a_0^2.
    }
We see that the orientifold gives these fields a mass
of order $m^2\sim g_sa_0^2$, which is the same
scale as generic low mass scales in this background.
Since this mass comes from the M\"obius strip, it is suppressed by a factor of
$g_s$ compared to other masses
so these open string moduli are still light, but not exponentially as
before, so hopefully they should not cause phenomenological problems.

\bigskip
\centerline{\bf Acknowledgements}

We would like to thank S. Kachru, S. Trivedi, and T. Volansky for
useful discussions. This work was supported in part by
the Israel-U.S. Binational Science Foundation, by the Israel Science
Foundation (grant number 1399/04), by the Braun-Roger-Siegl
foundation, by the European network HPRN-CT-2000-00122, by a grant
from the G.I.F., the German-Israeli Foundation for Scientific
Research and Development, by Minerva, by the Einstein center for
theoretical physics and by the Blumenstein foundation.

\appendix{A}{Low dimensional operators from supergravity analysis}

The Kaluza-Klein spectroscopy for the supergravity fields on
$AdS_5\times T^{1,1}$ was carried
out in \refs{\ceresole,\ceresoleb}. In this section we will review their
results for the dimensions of the corresponding operators as a function of
their $SU(2)\times SU(2)\times U(1)_r$ quantum numbers $j,l,r$. By
considering the group theoretic restrictions on these quantum numbers
for each multiplet, we will be able to find the operators with lowest
dimension.

In general we expand the fields in spherical harmonics on the $5$
dimensional compact space $T^{1,1}=\frac{SU(2)\times SU(2)}{U(1)}$,
with fields of different spins on the compact space
expanded using the corresponding $SO(5)$ harmonics. These harmonics
furnish representations of the isometry (global symmetry) group, $SU(2)\times
SU(2)\times U(1)_r$ in our case, but not all representations appear in
the expansion. The specific participating representations depend on
the Lorentz properties of the fields, but it turns out that all
representations satisfy that either both $SU(2)$ spins
$j$ and $l$ are integers or
both are half integers.

All resulting modes can be arranged into multiplets of the
${\cal N}=1$ $d=4$ superconformal algebra.
There are nine types of multiplets -- one graviton
multiplet, four gravitini and four vector multiplets \refs{
\ceresole,\ceresoleb}. For specific
values of the quantum numbers, some multiplets obey a shortening
condition and become semi-long, massless or chiral multiplets.

For the current analysis we are interested in operators that can be
turned on at some UV cut-off without breaking four dimensional
Lorentz invariance or supersymmetry. Hence the relevant multiplets
are only those with scalars as the highest component. These are only
the vector multiplets, either with generic values of the quantum
numbers or when they obey the condition for shortening to chiral
multiplets. The top components of these multiplets are related to
Kaluza-Klein modes of the warped $5$D metric on the $T^{1,1}$ $\hat
g_{ij}$, the field $\Phi_+$ defined in \fieldcomb, the three form
field $G_3$ and the axio-dilaton $\tau$.

The first vector multiplet (vector multiplet I in the notations of
\refs{\ceresole,\ceresoleb}) has a top component related to the
Kaluza-Klein modes of the warped $5$D metric on the $T^{1,1}$ $\hat
g_{ij}$, both when it is long and when it obeys the chiral shortening 
condition. The dimension
of this multiplet, defined as the dimension of the lowest component,
is given by
\eqn\vmienergy{
    \Delta =\sqrt{H(j,l,r)+4}-2,
    }
where
\eqn\ubiquitousH{
    H(j,l,r)\equiv 6(j(j+1)+l(l+1)-\frac{r^2}8),
    }
with $(j,l,r)$ the quantum numbers for the representation of the
$SU(2)\times SU(2)\times U(1)_r$ symmetry group. The lowest component
$b$, coming from a linear combination of the 5-form and the warp
factor $\Phi_-=\exp{4A}-\alpha$, is expanded in scalar harmonics that
satisfy that $r$ is even (odd) for $j,l$ integers (half integers) and
$|r|\le 2\,{\rm min}(j,l)$.

Small dimensions arise when $H$ is small. Due to
the $1/8$ factor in the third term, large values of $j$ and $l$
cannot be compensated by large values of $r$ and will give higher
values. It is then enough to look at small values for $j$ and $l$.
The lowest values and corresponding quantum numbers are written in
table 1. In addition, for each multiplet one can check whether it obeys some
shortening condition and what is the dimension of
the operator corresponding to the top component. The $j=l=r=0$
chiral operator can in fact be gauged away, so among the physical
scalar operators we are left with one
relevant operator and one marginal operator, and all others are irrelevant.

\vskip4pt
$$
\def\tbntry#1{\vbox to 20 pt{\vfill \hbox{#1}\vfill }}
\def\tbln#1#2#3#4#5#6#7{\hbox{\hbox to 25 pt{
                              \hfill\tbntry{$#1$}\hfill }
                              \vrule
                              \hbox to 25 pt{
                              \hfill\tbntry{$#2$}\hfill }
                              \vrule
                              \hbox to 25 pt{
                              \hfill\tbntry{$#3$}\hfill }
                              \vrule
                              \hbox to 30 pt{
                              \hfill\tbntry{$#4$}\hfill }
                              \vrule
                              \hbox to 25 pt{
                              \hfill\tbntry{$#5$}\hfill }
                              \vrule
                              \hbox to 50 pt{
                              \hfill\tbntry{#6}\hfill }
                              \vrule
                              \hbox to 25 pt{
                              \hfill\tbntry{$#7$}\hfill }
                              \vrule
                            }
                    \hrule
                    }
\hbox{\vrule
      \vbox{\hrule
            \tbln{j}{l}{|r|}{H}{\Delta}{Type}{\Delta_{top}}
            \hrule
            \hrule
            \tbln{0}{0}{0}{0}{0}{chiral}{1}
            \tbln{1/2}{1/2}{1}{8.25}{1.5}{chiral}{2.5}
            \tbln{0}{1}{0}{12}{2}{semilong}{-}
            \tbln{1}{0}{0}{12}{2}{semilong}{-}
            \tbln{1}{1}{2}{21}{3}{chiral}{4}
            \tbln{1}{1}{0}{24}{3.29}{none}{5.29}
            \tbln{1/2}{3/2}{1}{26.25}{3.5}{semilong}{-}
            \tbln{3/2}{1/2}{1}{26.25}{3.5}{semilong}{-}
        }
     }
$$
\centerline{ \hbox{{\bf Table 1:} {\it Lowest dimensional
operators from vector multiplet I.}}} \vskip 8pt

A similar analysis can be done for vector multiplet II for which the
top component is related to $\Phi_+$. This multiplet does not
satisfy any shortening condition. The dimension of the multiplet is
given by a similar expression
\eqn\vmiienergy{
    \Delta =\sqrt{H(j,l,r)+4}+4.
    }
In this case the top component is itself a mode of a ten dimensional
scalar field so the quantum numbers satisfy the same inequality as in
the previous case. The lowest dimensional operator has $H=0 \to
\Delta=6 \to \Delta_{top}=8$ which is already irrelevant. Some of the
low dimensional operators are described in table 2.

\vskip4pt
$$
\def\tbntry#1{\vbox to 20 pt{\vfill \hbox{#1}\vfill }}
\def\tbln#1#2#3#4#5#6#7{\hbox{\hbox to 25 pt{
                              \hfill\tbntry{$#1$}\hfill }
                              \vrule
                              \hbox to 25 pt{
                              \hfill\tbntry{$#2$}\hfill }
                              \vrule
                              \hbox to 25 pt{
                              \hfill\tbntry{$#3$}\hfill }
                              \vrule
                              \hbox to 30 pt{
                              \hfill\tbntry{$#4$}\hfill }
                              \vrule
                              \hbox to 25 pt{
                              \hfill\tbntry{$#5$}\hfill }
                              \vrule
                              \hbox to 50 pt{
                              \hfill\tbntry{#6}\hfill }
                              \vrule
                              \hbox to 25 pt{
                              \hfill\tbntry{$#7$}\hfill }
                              \vrule
                            }
                    \hrule
                    }
\hbox{\vrule
      \vbox{\hrule
            \tbln{j}{l}{|r|}{H}{\Delta}{Type}{\Delta_{top}}
            \hrule
            \hrule
            \tbln{0}{0}{0}{0}{6}{none}{8}
            \tbln{1/2}{1/2}{1}{8.25}{7.5}{none}{9.5}
            \tbln{0}{1}{0}{12}{8}{none}{10}
            \tbln{1}{0}{0}{12}{8}{none}{10}
            \tbln{1}{1}{2}{21}{9}{none}{11}
            \tbln{1}{1}{0}{24}{9.29}{none}{11.29}
            \tbln{1/2}{3/2}{1}{26.25}{9.5}{none}{11.5}
            \tbln{3/2}{1/2}{1}{26.25}{9.5}{none}{11.5}
        }
     }
$$
\centerline{ \hbox{{\bf Table 2:} {\it Lowest dimensional
operators from vector multiplet II.}}} \vskip 8pt

For the vector multiplet III, the top component (whether or not the
multiplet obeys a shortening condition)
is related to the three form field $G_3$, and the dimension
of the multiplet is
\eqn\vmiiienergy{
    \Delta =\sqrt{H(j,l,r+2)+4}+1.
    }
For this multiplet none of the fields are expanded in scalar
harmonics. Instead, the top component $a$ (which is the same for both
the long and chiral multiplets) originating from the ten dimensional
two-form potential is expanded using the two-form harmonics. For
these harmonics we again have that $r$ is even (odd) for $j,l$
integers (half integers), but now the restriction on the quantum
numbers is $|r|\le 2\,{\rm min}(j,l)+2$.

In this case we also have non trivial restrictions from the
unitarity bounds
\eqn\unitbound{
    2-\Delta \le\frac32r\le \Delta -2.
    }
The possible quantum numbers are given in table 3.

\vskip4pt
$$
\def\tbntry#1{\vbox to 20 pt{\vfill \hbox{#1}\vfill }}
\def\tbln#1#2#3#4#5#6#7{\hbox{\hbox to 25 pt{
                              \hfill\tbntry{$#1$}\hfill }
                              \vrule
                              \hbox to 25 pt{
                              \hfill\tbntry{$#2$}\hfill }
                              \vrule
                              \hbox to 25 pt{
                              \hfill\tbntry{$#3$}\hfill }
                              \vrule
                              \hbox to 30 pt{
                              \hfill\tbntry{$#4$}\hfill }
                              \vrule
                              \hbox to 25 pt{
                              \hfill\tbntry{$#5$}\hfill }
                              \vrule
                              \hbox to 50 pt{
                              \hfill\tbntry{#6}\hfill }
                              \vrule
                              \hbox to 25 pt{
                              \hfill\tbntry{$#7$}\hfill }
                              \vrule
                            }
                    \hrule
                    }
\hbox{\vrule
      \vbox{\hrule
            \tbln{j}{l}{r}{H}{\Delta}{Type}{\Delta_{top}}
            \hrule
            \hrule
            \tbln{0}{0}{0}{-3}{2}{chiral}{4}
            \tbln{1/2}{1/2}{1}{2.25}{3.5}{chiral}{5.5}
            \tbln{1/2}{1/2}{-1}{8.25}{4.5}{none}{6.5}
            \tbln{0}{1}{0}{9}{4.61}{none}{6.61}
            \tbln{1}{0}{0}{9}{4.61}{none}{6.61}
        }
     }
$$
\centerline{ \hbox{{\bf Table 3:} {\it Lowest dimensional
operators from vector multiplet III.}}} \vskip 8pt

Finally, vector multiplet IV has dimension given by
\eqn\vmivenergy{
    \Delta =\sqrt{H(j,l,r-2)+4}+1.
    }
For long multiplets the top component is related to the three form
field $G_3$ while for multiplets satisfying a chiral shortening
condition the top component is related to the axio-dilaton $\tau$.
This last mode appear in all vector multiplets of this type 
(though it is not always the top component)
and it is expanded in scalar harmonics. Its r-charge equal to
$r-2$ so the quantum numbers obey $|r-2|\le 2\,{\rm min}(j,l)$. Since
the dimension depends only on $r-2$ we get a similar table to that of
vector multiplet I but with the dimensions shifted:

\vskip4pt
$$
\def\tbntry#1{\vbox to 20 pt{\vfill \hbox{#1}\vfill }}
\def\tbln#1#2#3#4#5#6#7{\hbox{\hbox to 25 pt{
                              \hfill\tbntry{$#1$}\hfill }
                              \vrule
                              \hbox to 25 pt{
                              \hfill\tbntry{$#2$}\hfill }
                              \vrule
                              \hbox to 25 pt{
                              \hfill\tbntry{$#3$}\hfill }
                              \vrule
                              \hbox to 30 pt{
                              \hfill\tbntry{$#4$}\hfill }
                              \vrule
                              \hbox to 25 pt{
                              \hfill\tbntry{$#5$}\hfill }
                              \vrule
                              \hbox to 50 pt{
                              \hfill\tbntry{#6}\hfill }
                              \vrule
                              \hbox to 25 pt{
                              \hfill\tbntry{$#7$}\hfill }
                              \vrule
                            }
                    \hrule
                    }
\hbox{\vrule
      \vbox{\hrule
            \tbln{j}{l}{r}{H}{E_0}{Type}{\Delta_{top}}
            \hrule
            \hrule
            \tbln{0}{0}{0}{0}{3}{chiral}{4}
            \tbln{1/2}{1/2}{1}{8.25}{4.5}{chiral}{5.5}
            \tbln{0}{1}{0}{12}{5}{semilong}{-}
            \tbln{1}{0}{0}{12}{5}{semilong}{-}
            \tbln{1}{1}{2}{21}{6}{chiral}{7}
            \tbln{1}{1}{0}{24}{6.29}{none}{8.29}
            \tbln{1/2}{3/2}{1}{26.25}{6.5}{semilong}{-}
            \tbln{3/2}{1/2}{1}{26.25}{6.5}{semilong}{-}
        }
     }
$$
\centerline{ \hbox{\rm {\bf Table 4:} {\it Lowest dimensional
operators from vector multiplet IV.}}} 
\vskip 8pt
\break

To summarize this appendix we include a table of the lowest dimension
operators found above and their dimensions, as well as the form of the
corresponding operators in the field theory when it is known.
The first irrelevant operator is the top component of a
long multiplet and contributes to the K\"ahler potential, but its exact
form is not known.

\vskip4pt
$$
\def\tbntry#1{\vbox to 20 pt{\vfill \hbox{#1}\vfill }}
\def\tbln#1#2#3#4#5#6#7{\hbox{\hbox to 25 pt{
                              \hfill\tbntry{$#1$}\hfill }
                              \vrule
                              \hbox to 25 pt{
                              \hfill\tbntry{$#2$}\hfill }
                              \vrule
                              \hbox to 25 pt{
                              \hfill\tbntry{$#3$}\hfill }
                              \vrule
                              \hbox to 30 pt{
                              \hfill\tbntry{$#4$}\hfill }
                              \vrule
                              \hbox to 50 pt{
                              \hfill\tbntry{#5}\hfill }
                              \vrule
                              \hbox to 50 pt{
                              \hfill\tbntry{#6}\hfill }
                              \vrule
                              \hbox to 130 pt{
                              \hfill\tbntry{$#7$}\hfill }
                              \vrule
                            }
                    \hrule
                    }
\hbox{\vrule
      \vbox{\hrule
            \tbln{\Delta}{j}{l}{|r|}{Multiplet}{Type}{{\rm Operator}}
            \hrule
            \hrule
            \tbln{2.5}{1/2}{1/2}{1}{I}{chiral}{S_1=\int\!\! d^2\theta\ \Tr(AB)}
            \tbln{4}{1}{1}{2}{I}{chiral}{S_2=\int\!\! d^2\theta\ \Tr[(AB)^2]}
            \tbln{4}{0}{0}{0}{IV}{chiral}{\Phi_0=\int\!\! d^2\theta\ \Tr(W_1^2+W_2^2)}
            \tbln{4}{0}{0}{0}{III}{chiral}{\Psi_0=\int\!\! d^2\theta\ \Tr(W_1^2-W_2^2)}
            \tbln{5.29}{1}{1}{0}{I}{long}{\CO_1=\int\!\! d^4\theta\ (?)}
        }
     }
$$
\centerline{ \hbox{{\bf Table 5:} {\it Lowest dimensional
operators.}}} \vskip 8pt

\appendix{B}{The moduli space of the deformed theory}

In this appendix we consider deforming the superpotential of the
Klebanov-Witten theory \kw\ by the relevant and marginal operators $S_1$
and $S_2$ given in \relop, \margop. The resulting moduli space is
analyzed and shown to be of lower dimension than the original
symmetric product of $N$ copies of the conifold. In the dual gravity
description this must be due to a deformation exerting a force on the
D-branes. Such a deformation is forbidden by the equations of motion
in our construction, so we conclude that these operators are not turned
on. For the case of a Klebanov-Strassler background
the conclusion should remain the same.

In the $SU(N)\times SU(N)$ theory with superpotential
$W = h \epsilon_{ik} \epsilon_{jl} \tr(A^i B^j A^k B^l)$
the F-term equations require
that the chiral fields $A_i, B_i\quad i=1,2$ will commute, so that they
can be simultaneously diagonalized by gauge transformations. The
D-term equations then lead to the general solution being
$N$ copies of the conifold.
This branch describes D-branes moving separately on the 6
dimensional geometry. Such a branch must also exist for the
deformed theory due to the no force condition on the D-branes, and we expect
that
the subspace of diagonal matrices that solve the F-term and D-term
equations should give us the $N$'th symmetric product of the
deformed 6 dimensional geometry.

For diagonal matrices the equations for the $N\times N$ matrices
become decoupled so we can consider them as $N$ identical equations for
single fields. In this case the moduli space will be the solution to the
F-term equations divided by the complexification of the $U(1)$ gauge
group, and the original superpotential can be ignored. Since all
fields are charged under the $U(1)$ we get
\eqn\msdim{
    {\rm Dim}({\rm moduli\ space})={\rm Dim}({\rm F-term\ solutions})-2.
    }
Since the dimension should be six, we get that the solutions to the F-term
equations should form an $8$ dimensional space.

We begin by considering the relevant chiral operator $S_1$. The
general deformation of the superpotential is given by:
\eqn\abspot{
    \Delta W=\lambda_{ij}\Tr(A_iB_j),
    }
where $\lambda_{ij}$ is constant matrix. The F-term equations for
$1\times 1$ scalars are:
\eqn\abfterm{
    \lambda\cdot B = 0,\qquad \lambda^t\cdot A=0,
    }
where we consider $A,B$ as 2-vectors and $\lambda$ as a $2\times
2$ matrix. For $\det(\lambda)\ne0$ there is no solution to the
system of equations.
For $\det(\lambda)=0,\ \lambda\ne0$
(so ${\rm rank}(\lambda)=1$), there is a 2 dimensional space of complex
solutions so the moduli space is 2 dimensional. Only for
$\lambda=0$ we get a moduli space large enough for describing free
D-branes.

We now add also the marginal operator $S_2$. The deformed
superpotential is
\eqn\ababspot{
    \Delta
    W=\lambda_{ij}\Tr(A_iB_j)+\frac12\sigma_{ijkl}\Tr(A_iB_jA_kB_l),
    }
with $\sigma_{ijkl}=\sigma_{kjil}=\sigma_{ilkj}$. The symmetry
condition for the indices comes from the fact that the chiral
marginal operator is the $j=l=1$ combination of the four
fields. From the cyclicity of the trace we also get
$\sigma_{ijkl}=\sigma_{klij}$.

The F-term equations for $1\times 1$ scalars are now
\eqn\ababfterm{\eqalign{
    \lambda_{ij}B_j+\frac12\sigma_{ijkl}B_jA_kB_l+
\frac12\sigma_{kjil}B_jA_kB_l&=\lambda_{ij}B_j+\sigma_{ijkl}B_jA_kB_l=0,\cr
    \lambda_{ji}A_j+\sigma_{jikl}A_kB_lA_j&=0,
    }}
where we used the symmetries of $\sigma$. There are 4 complex fields
in the equations so the maximal dimension for the space of solutions
is 8. If the solutions indeed form an 8 dimensional space then for a
generic solution any change $A_i\to A_i+\delta A_i$ and $B_i\to
B_i+\delta B_i$ will result in a new solution. Taking the first
equation and shifting only the $A_i$ fields we find for any $\delta
A_i$
\eqn\dafterm{\eqalign{
    \lambda_{ij}B_j+&\sigma_{ijkl}B_jA_kB_l+\sigma_{ijkl}B_j\delta A_kB_l=0\cr
    &\Rightarrow \sigma_{ijkl}B_j\delta A_kB_l=0\cr
    &\quad\!\! \Rightarrow \sigma_{ijkl}B_jB_l=0.
    }}
Again this holds for all solutions so we can now shift $B_i$ to
find
\eqn\dadbfterm{\eqalign{
    \sigma_{ijkl}B_jB_l+&\sigma_{ijkl}\delta B_jB_l+\sigma_{ijkl}B_j\delta B_l+\sigma_{ijkl}\delta B_j\delta B_l=0\cr
    &\Rightarrow (\sigma_{ijkl}+\sigma_{ilkj})B_j\delta B_l=0\cr
    &\Rightarrow \sigma_{ijkl}+\sigma_{ilkj}=0\cr
    &\Rightarrow \sigma_{ijkl}=0.
    }}
Where in the last step we used the symmetry properties of
$\sigma_{ijkl}$. Hence we are left with only the relevant
deformation which also vanishes by the previous argument.

This argument can be generalized to any higher deformation of this
form. Consider the deformation $\Tr(A_{i_1}B_{j_1}\cdots
A_{i_n}B_{j_n})$. Since the $SU(2)\times SU(2)$ representation is
$j=l=\frac n2$ the coefficient of this term is symmetric under
exchange of the $j$ indices and under exchange of the $i$ indices. We
can then carry out a similar argument, where we take at each step
another derivative with respect to $A$ or $B$. Due to the symmetry
property, each time we will get the same coefficient with one lower
power of the fields. After $2n$ steps we will be left with only this
term and no lower terms, and will arrive to the conclusion that the
coefficient must vanish.

We conclude that turning on these types of deformations the diagonal
branch of the moduli space cannot have 6 dimensions. Since the
geometry discussed in section 2 accommodates D-branes on a 6
dimensional space, these operators are not turned on by deforming the
Klebanov-Strassler solution to a compact Calabi-Yau. In particular
the relevant and marginal deformations ($n=1$ and $n=2$) are not
turned on.

\listrefs

\end